\documentclass[11pt,preprint,showpacs,preprintnumbers,amsmath,amssymb]{revtex4}

\usepackage{exscale}
\usepackage{relsize}
\usepackage{graphicx}% Include figure files
\usepackage{dcolumn}% Align table columns on decimal point
\usepackage{bm}% bold math
\usepackage{multirow}
\usepackage{CJK}
\usepackage{diagbox}
\usepackage{subfigure}
\usepackage{epstopdf}

\begin{document}

\title{Studying the bound state of the $B\bar{K}$ system in the Bethe-Salpeter formalism}

\author{Zhen-Yang Wang \footnote{e-mail: wangzhenyang@nbu.edu.cn}}
\affiliation{\scriptsize{Physics Department, Ningbo University, Zhejiang 315211, China}}

\author{Jing-Juan Qi \footnote{e-mail: qijj@mail.bnu.edu.cn}}
\affiliation{\scriptsize{College of Nuclear Science and Technology, Beijing Normal University, Beijing 100875, China}}

\author{Xin-Heng Guo \footnote{Corresponding author, e-mail: xhguo@bnu.edu.cn}}
\affiliation{\scriptsize{College of Nuclear Science and Technology, Beijing Normal University, Beijing 100875, China}}

\date{\today}

\begin{abstract}
In this work, we study the $B\bar{K}$ molecule in the Bethe-Salpeter (BS) equation approach. With the kernel containing one-particle-exchange diagrams and introducing two different form factors (monopole form factor and dipole form factor) in the vertex, we solve the BS equation numerically in the covariant instantaneous approximation. We investigate the isoscalar and isovector $B\bar{K}$ systems, and we find $X(5568)$ cannot be a $B\bar{K}$ molecule.
\end{abstract}


\maketitle
\section{Introduction}
The physics of exotic multiquark states has been a subject of intense interest in recent years. One reason for this is that with the experimental data are being accumulated on charmonium-like $XYZ$ states and $P_c$ pentaquark states (see the review papers \cite{Guo:2017jvc,Esposito:2016noz,Chen:2016spr} for details) and more and more experimental data will be found in near future.

In 2016, the D0 Collaboration announced a new enhancement structure $X(5568)$ with the statistical significance of 5.1$\sigma$ in the $B_s^0\pi^{\pm}$ invariant mass spectrum, which has the mass $5567.8\pm2.9\mathrm{(stat)}^{+0.9}_{-1.9}$(syst) MeV and width $\Gamma=21.9\pm6.4\mathrm{(stat)}^{+5.0}_{-2.5}$(syst) MeV \cite{D0:2016mwd}. The observed channel indicates the isospin of the $X(5568)$ is 1 and if it decays into $B^0\pi^\pm$ via a S-wave, the quantum numbers of the $X(5568)$ should be $J^P = 0^+$. Subsequent analyses by the LHCb \cite{Aaij:2016iev}, CMS \cite{Sirunyan:2017ofq}, and ATLAS \cite{Aaboud:2018hgx} Collaborations have not found evidence for the $X(5568)$ in proton-proton interactions at $\sqrt{s}$= 7 and 8 TeV. The CDF Collaboration has recently reported no evidence for $X(5568)$ in proton-antiproton collisions at $\sqrt{s}$ = 1.96 TeV \cite{Aaltonen:2017voc} with different kinematic. Recently, the D0 Collaboration reported a further evidence about this state in the decay of $B$ with a significance of 6.7$\sigma$ \cite{Abazov:2017poh} which is consistent with their previous measurement in the hadronic decay mode \cite{D0:2016mwd}. Therefore, the experimental status of the $X(5568)$ resonance remains unclear and controversial.

No matter whether the structure exists or not, it has been attracting a lot of attention from both experimental and theoretical sides. Many theoretical groups have studied possible ways to explain $X(5568)$ as a tetraquark state, a molecular state, etc. within various models, and they obtained different results. In Refs. \cite{Agaev:2016mjb,Zanetti:2016wjn,Chen:2016mqt,Agaev:2016ijz,Dias:2016dme,Wang:2016wkj,Tang:2016pcf,Agaev:2016ifn,Zhang:2017xwc}, the authors based on QCD sum rules obtained the mass and/or decay width which are in agreement with the experimental data. In Refs. \cite{Xiao:2016mho,Chen:2016ypj}, the authors showed that $X(5568)$ or $X(5616)$ could not be assigned to be an $B\bar{K}$ or $B^*\bar{K}$ molecular state. $X(5568)$ is also disfavored as a $P$-wave coupled-channel scattering molecule involving the states $B_s\pi$, $B_s^*\pi$, $B\bar{K}$ and $B^*\bar{K}$ in Ref. \cite{Kang:2016zmv}. The  authors of Ref. \cite{Lu:2016kxm} pointed out the $B_s\pi$ and $B\bar{K}$ interactions were weak and $X(5568)$ could not be a $S$-wave $B_s\pi$ and $B\bar{K}$ molecular state. Based on the lattice QCD, there is no candidate for $X(5568)$ with $J^P=0^+$ \cite{Lang:2016jpk}. The authors found that threshold, cusp, molecular, and tetraquark models were all unfavoured for $X(5568)$ \cite{Burns:2016gvy}. $X(5568)$ as $B\bar{K}$ molecule and diquark-diquark model are considered in Ref. \cite{Agaev:2016urs} using QCD two-point and light-cone sum rules, and their results strengthens the diquark-antidiquark picture for the $X(5568)$ state rather than a meson molecule structure. But the authors of Ref. \cite{Albaladejo:2016eps} found that the $X(5568)$ signal can be reproduced by using $B_s\pi - B\bar{K}$ coupled channel analysis, if the corresponding cutoff value was larger than a natural value $\Lambda$ $\sim$ 1 GeV. In Ref. \cite{Yang:2016sws}, the authors demonstrated that $X(5568)$ could be a kinematic reflection and explained the absence of $X(5568)$ in LHCb and CMS Collaborations. Based on the quark model, $X(5568)$ could exist as a mixture of a tetraquark and hadronic molecule \cite{Ke:2018stp}.

By this chance, we will systematically study the $B\bar{K}$ molecular state in the BS equation approach. We investigate the $S$-wave $B\bar{K}$ systems with both isospin $I =0, 1$ are consider. We will vary
 $E_b(E_b=E-M_{B}-M_K)$ in a much wider range and search for all the possible solutions. In this process, we naturally check whether $X(5568)$ can exist as $S$-wave $B\bar{K}$ molecular state, or not.

The remainder of this paper is organized as follows. In Sec. \ref{sect-BS-PP}, we discuss the BS equation for two pseudoscalar mesons and establish the one-dimensional BS function for this system. The numerical results of the $B\bar{K}$ systems are presented in Sec. \ref{Num}. In the last section, we give a summary and some discussions.

\section{the bethe-salpeter formalism for $B\bar{K}$ system}
\label{sect-BS-PP}
In this section, we will review the general formalism of the BS equation and establish the BS equation for the system of two pseudoscalar mesons. Let us start by defining the BS wave function for the bound state $|P\rangle$ as the following:
\begin{equation}
  \chi\left(x_1,x_2,P\right) = \langle0|TB(x_1)\bar{K}(x_2)|P\rangle,
\end{equation}
where $B(x_1)$ and $\bar{K}(x_2)$ are the field operators of the $B$ and $\bar{K}$ mesons at space coordinates $x_1$ and $x_2$, respectively, $P$ denotes the total momentum of the bound state with mass $M$ and velocity $v$. The BS wave function in momentum space is defined as
\begin{equation}\label{PP-momentum-BS-function}
 \chi_P(x_1,x_2,P) = e^{-iPX}\int\frac{d^4p}{(2\pi)^4}e^{-ipx}\chi_P(p),
\end{equation}
where $p$ represents the relative momentum of the two constituents and $p= \lambda_2 p_1-\lambda_1 p_2$ (or $p_1=\lambda_1P+p$,\quad $p_2=\lambda_2P-p$). The relative coordinate $x$ and the center-of-mass coordinate $X$ are defined by
\begin{equation}\label{co}
  X = \lambda_1x_1 + \lambda_2x_2, \quad x = x_1 - x_2,
\end{equation}
or inversely,
\begin{equation}
  x_1 = X + \lambda_2x,   \quad   x_2 = X - \lambda_1x,
\end{equation}
where $\lambda_1 = m_{B}/(m_{B} + m_K)$ and $\lambda_2 = m_K/(m_{B} + m_K)$, $m_{B}$ and $m_K$ are the masses of $B$ and $K$ mesons.

It can be shown that the BS wave function of $B\bar{K}$ bound state satisfies the following BS equation \cite{Lurie}:
\begin{equation}\label{BS-equation}
  \chi_{P}(p)=S_{B}(p_1)\int\frac{d^4q}{(2\pi)^4}K(P,p,q)\chi_{P}(q)S_{\bar{K}}(p_2),
\end{equation}
where $S_{B}$ and $S_{\bar{K}}(p_2)$ are the propagators of $B$ and $\bar{K}$, respectively, and $K(P,p,q)$ is the kernel, which is defined as the sum of all the two particle irreducible diagrams with respect to $B$ and $\bar{K}$ mesons. For convenience, in the following we use the variables $p_l (=p\cdot v)$ and $p_t(=p- p_lv)$ to be the longitudinal and transverse projections of the relative momentum ($p$) along the bound state momentum ($P$). Then, the propagator of $B$ mesons can be expressed as
\begin{equation}\label{B-propagator}
  S_B(\lambda_1P+p)=\frac{i}{\left(\lambda_1M+p_l\right)^2-\omega_1^2+i\epsilon},
\end{equation}
and the the propagator of the $\bar{K}$ is
\begin{equation}\label{K-propagator}
  S_K(\lambda_2P-p)=\frac{i}{\left(\lambda_2M-p_l\right)^2-\omega_2^2+i\epsilon},
\end{equation}
where $\omega_{1(2)} = \sqrt{m_{B(K)}^2+p_t^2}$ (we have defined $p_t^2=-p_t\cdot p_t$).

As discussed in the introduction, we will study the $S$-wave bound state of $B\bar{K}$ system. The field doublets $(B^{+},B^{-})$, $(B^{0},\bar{B}^{0})$, $(K^+,K^-)$ and $(K^0,\bar{K}^0)$ have the following expansions in momentum space:
\begin{equation}
  \begin{split}
    B_1(x) &= \int\frac{d^3p}{(2\pi)^3\sqrt{2E_{B}^{\pm}}}\left(a_{B^{+}}e^{-ipx}+a_{B^{-}}^\dag e^{ipx}\right), \\
    B_2(x) &= \int\frac{d^3p}{(2\pi)^3\sqrt{2E_{B}^0}}\left(a_{B^{0}}e^{-ipx}+a_{\bar{B}^{0}}^\dag e^{ipx}\right), \\
    K_1(x) &= \int\frac{d^3p}{(2\pi)^3\sqrt{2E_K^{\pm}}}\left(a_{K^-}e^{-ipx}+a_{\bar{K}^+}^\dag e^{ipx}\right), \\
    K_2(x) &= \int\frac{d^3p}{(2\pi)^3\sqrt{2E_K^0}}\left(a_{K^0}e^{-ipx}+a_{\bar{K}^0}^\dag e^{ipx}\right), \\
  \end{split}
\end{equation}
where $E_{B(K)}=\sqrt{p^2_{1(2)}+m^2_{B(K)}}$ is the energy of the particle.

The isospin of $B\bar{K}$ can be 0 or 1 for $B\bar{K}$ system, and the flavor wave function for the isoscalar bound state can be written as
\begin{equation}
  |P\rangle_{0,0} = \frac{1}{\sqrt{2}}|B^{+}K^- + B^{0}\bar{K}^0\rangle,
\end{equation}
and the flavor wave functions of the isovector states for $B\bar{K}$ system are
\begin{equation}
\begin{split}
  &|P\rangle_{1,1} = |B^{+}\bar{K}^0\rangle,\quad\quad |P\rangle_{1,0} = \frac{1}{\sqrt{2}}|B^{+}K^- - B^{0}\bar{K}^0\rangle,\quad\quad|P\rangle_{1,-1} = |B^{0}K^-\rangle.
\end{split}
\end{equation}

Let us now project the bound states on the field operators $B_1(x)$, $B_2(x)$, $K_1(x)$ and $K_2(x)$. Then we have
\begin{equation}
  \langle0|T{B_i(x_1)K_j(x_2)}|P\rangle_{I,I_3}=C_{(I,I_3)}^{ij}\chi_P^{(\mu) I}\left(x_1,x_2\right),
\end{equation}
where $\chi_P^{I}$ is the common BS wave function for the bound state with isospin $I$ which depends only on $I$ but not $I_3$ of the state $|P\rangle_{I,I_3}$. The isospin coefficients $C_{(I,I_3)}^{ij}$ for the isoscalar state are
\begin{equation}\label{isos-isospin-coef}
C_{(0,0)}^{11}=C_{(0,0)}^{22}=1/\sqrt{2}, \quad\quad \mathrm{else}=0,
\end{equation}
and for the isovector states we have
\begin{equation}\label{isov-isospin-coef}
C_{(1,1)}^{12}=C_{(1,-1)}^{21}=1, \quad\quad C_{(1,0)}^{11} =-C_{(1,0)}^{22}=1/\sqrt{2}, \quad\quad \mathrm{else}=0.
\end{equation}

Now considering the kernel, Eq. (\ref{BS-equation}) can be written down schematically,
\begin{equation}
  C_{(I,I_3)}^{ij}\chi_P^{(\mu) I}(p) = S_{B}(\lambda_1P+p) \int\frac{d^4q}{(2\pi)^4}K^{ij,lk}\left(P,p,q\right) C_{(I,I_3)}^{lk}\chi_P^{I}(q)S_{\bar{K}}(\lambda_2P-p),
\end{equation}
then, from Eq. (\ref{isos-isospin-coef}), for the isoscalar case, we have (take
$ij = 11$ as an example)
\begin{equation}
  \chi_P^{0}(p) = S_{B}(\lambda_1P+p) \int\frac{d^4q}{(2\pi)^4}\left[K^{11,11}+K^{11,22}\right]\chi_P^{0}(q)S_{\bar{K}}(\lambda_2P-p).
\end{equation}
Similarly, for the isovector case, taking the $I_3 = 0$ component as an example, we have\begin{equation}
  \chi_P^{1}(p) = S_{B}(\lambda_1P+p) \int\frac{d^4q}{(2\pi)^4}\left[K^{11,11}-K^{11,22}\right]\chi_P^{1}(q)S_{\bar{K}}(\lambda_2P-p).
\end{equation}

In the BS equation approach, the interaction between $B$ and $\bar{K}$ mesons can be due to the light vector-meson ($\rho$ and $\omega$) exchanges. The corresponding effective Lagrangians describing the couplings of $BB\rho(\omega)$ \cite{He:2014nya,Feng:2011zzb} and $KK\rho(\omega)$ \cite{Feng:2012zzf,Chen:2017xat} are
\begin{equation}
\begin{split}\label{BK-Lagrangian}
  \mathcal{L}_{BB\mathbb{V}} &= -ig_{BB\mathbb{V}}B_a^\dag\overleftrightarrow{\partial}B_b\mathbb{V}_{ba}^\mu,\\
   \mathcal{L}_{KK\rho} &= ig_{KK\rho}\left[K^\dag\vec{\tau}(\partial_\mu K)-(\partial_\mu K^\dag)\vec{\tau}K\right]\cdot\vec{\rho}^\mu,\\
  \mathcal{L}_{KK\omega} &= ig_{KK\omega}\left[K^\dag(\partial_\mu K)-(\partial_\mu K^\dag)K\right]\omega^\mu,\\
  \end{split}
\end{equation}
where the nonet vector meson matrix read as
\begin{eqnarray}
\mathbb{V}&=&\left(\begin{array}{ccc}
\frac{\rho^{0}}{\sqrt{2}}+\frac{\omega}{\sqrt{2}}&\rho^{+}&K^{*+}\\
\rho^{-}&-\frac{\rho^{0}}{\sqrt{2}}+\frac{\omega}{\sqrt{2}}&
K^{*0}\\
K^{*-} &\bar{K}^{*0}&\phi
\end{array}\right).\label{vector}
\end{eqnarray}
In addition, the coupling constants involved in Eq. (\ref{BK-Lagrangian}) are taken as $g_{BB\mathbb{V}}=\frac{\beta g_v}{\sqrt{2}}$ with $g_v = 5.8$, $\beta = 0.9$, while the coupling constants $g_{KK\mathbb{V}}$ satisfy the relations $g_{KK\rho}=g_{KK\omega}=g_{\rho\pi\pi}/2$ in the $SU(3)_f$ limit, and $g_{\rho\pi\pi}\simeq m_\rho/f_\pi\simeq 5.8$ \cite{Feng:2012zzf}.

From the above observations, at the tree level, in the $t$-channel the kernel for the BS equation of the interaction between $B$ and $\bar{K}$ in the so-called lader approximation is taken to has the following form:
\begin{equation}\label{PP-kernel}
\begin{split}
  K_{V}(P,p,q)&=(2\pi)^4\delta^4(q_1+q_2-p_1-p_2)c_I g_{BBV}g_{KKV}\left(p_1+q_1\right)_\mu(p_2+q_2)_\nu\Delta^{\mu\nu}\left(k,m_V\right),\\
\end{split}
\end{equation}
where $m_V$ represent the masses of the exchanged light vector meson $\rho$ and $\omega$ , $c_I$ is the isospin coefficient: $c_0 =3, 1$ and $c_1 =1, 1$ for $\rho$, $\omega$, and $\Delta^{\mu\nu}$ represents the propagator for vector meson.

In order to manipulate the off shell effect of the exchanged mesons $\rho$ and $\omega$, and finite size effect of the interacting hadrons, we introduce a form factor $\mathcal{F}(k^2)$ at each vertex. Generally, the form factor has the monopole form and dipole form as shown in Ref. \cite{Chen:2017vai}
\begin{equation}\label{monopole-form-factor}
\mathcal{F}_M(k^2)=\frac{\Lambda^2-m^2}{\Lambda^2-k^2},
\end{equation}
\begin{equation}\label{dipole-form-factor}
\mathcal{F}_D(k^2)=\frac{(\Lambda^2-m^2)^2}{(\Lambda^2-k^2)^2},
\end{equation}
where $\Lambda$, $m$ and $k$ represent the cutoff parameter, mass of the exchanged meson and momentum of the exchanged meson, respectively. These two kinds of form factors are normalized at the on shell momentum of $k^2=m^2$. On the other hand, if $k^2$ were taken to be infinitely large ($-\infty$), the form factors, which can be expressed as the overlap integral of the wave functions of the hadrons at the vertex, would approach zero.

For the $B\bar{K}$ system, substituting Eqs. (\ref{B-propagator}), (\ref{K-propagator}), (\ref{PP-kernel}) and aforementioned form factors Eqs. (\ref{monopole-form-factor}) and (\ref{dipole-form-factor}) into Eq. (\ref{BS-equation}) and using the so-called covariant instantaneous approximation \cite{Guo:1996jj}, $p_l=q_l$ (which ensures that the BS equation is still covariant after this approximation). Then one obtains the expression
\begin{equation}\label{PP-expend-BS-equation}
  \begin{split}
    \chi\left(p\right) =&\frac{ic_Ig_{BBV}g_{KKV}}{[(\lambda_1M+p_l)^2-\omega_1^2+i\epsilon][(\lambda_2M-p_l)^2-\omega_2^2+i\epsilon]}\int\frac{d^4q}{(2\pi)^4}\\
    &\frac{4(\lambda_1M+p_l)(\lambda_2M-p_l)+(p_t+q_t)^2+(p_t^2-q_t^2)^2/m_V^2}{-(p_t-q_t)^2-m_V^2}\mathcal{F}^2(k_t)\chi\left(q\right).\\
  \end{split}
\end{equation}

In Eq. (\ref{PP-expend-BS-equation}) there are poles in $-\lambda_1M-\omega_1-i\epsilon$, $-\lambda_1M+\omega_1-i\epsilon$, $\lambda_2M+\omega_2-i\epsilon$, and $\lambda_2M-\omega_2+i\epsilon$. By choosing the appropriate contour, we integrate over $p_l$ on both sides of Eq. (\ref{PP-expend-BS-equation}) in the rest frame, we will obtain the following equation
%\begin{small}
\begin{equation}\label{int-ql-BS-equation}
\begin{split}
  \tilde{\chi}(p_t)=&\frac{c_I g_{BBV}g_{KKV}}{2(M+\omega_1-\omega_2)}\int\frac{d^3p_t}{(2\pi)^3}\Bigg[\frac{-4\omega_1(M
  +\omega_1)+(p_t+q_t)^2+(p_t^2-q_t^2)^2/m_V^2}{\omega_1(M+\omega_1+\omega_2)\left[-(p_t-q_t)^2-m_V^2\right]}\\
  &-\frac{4\omega_2(M-\omega_2)+(p_t+q_t)^2+(p_t^2-q_t^2)^2/m_V^2} {\omega_2(M-\omega_1-\omega_2)\left[-(p_t-q_t)^2-m_V^2\right]}\Bigg]\mathcal{F}^2(k_t)\tilde{\chi}(q_t),\\
  \end{split}
\end{equation}
%\end{small}
where $\tilde{\chi}(p_t)=\int dp_l\chi(p)$.

\section{Numerical results}
\label{Num}
In this part, we will solve the BS equation numerically and study whether the S-wave $B\bar{K}$ bound state exists or not. It can be seen from Eq. (\ref{int-ql-BS-equation}) that there is only one free
parameter in our model, the cutoff $\Lambda$, it can not be uniquely determined, and various forms and cutoff $\Lambda$ are chosen phenomenologically. It contains the information about the nonpoint
interaction due to the structures of hadrons. The value of $\Lambda$ is near 1 GeV which is the typical scale of nonperturbative QCD interaction. In this work, we shall treat the cutoff $\Lambda$ in the form factors as a parameter varying in a much wider range 0.8-4.8 GeV, in which we will try to search for all the possible solutions of the $B\bar{K}$ bound states. For each pair of trial values of the cutoff $\Lambda$ and the binding energy $E_b$ of the $B\bar{K}$ system (which is defined as $E_b = E-m_1-m_2$), we will obtain all the eigenvalues of this eigenvalue equation. The eigenvalue closest to 1.0 for a pair of $\Lambda$ and $E_b$ will be selected out and called ¡®¡®the trial eigenvalue.¡¯¡¯ Fixing a value of the cutoff $\Lambda$ and varying the binding energy $E_b$ (from 0 to -220 MeV) we will obtain a series of the trial eigenvalues.

Since the BS wave function for the ground state is in fact rotationally invariant, $\tilde{\chi}(p_t)$ depends only on $|p_t|$, Generally, $|p_t|$ varies from 0 to +$\infty$ and $\tilde{\chi}(p_t)$ would decrease to zero when $|p_t|\rightarrow + \infty$. We replace $|p_t|$ by the variable, $t$:
\begin{equation}
|p_t|=\epsilon+w\log\left[1+y\frac{1+t}{1-t}\right],
\end{equation}
where $\epsilon$ is a parameter introduced to avoid divergence in numerical calculations, $w$ and $y$ are parameters used in controlling the slope of wave functions and finding the proper solutions for these functions,  $t$ varies from -1 to 1. We then discretize Eq. (\ref{int-ql-BS-equation}) into $n$ pieces ($n$ is large enough) through the Gauss quadrature rule. The BS wave function can be written as $n$-dimension vectors, $\tilde{\chi}(p_t)$
. The coupled integral equation becomes a matrix equation $\tilde{\chi}(|p_t|(n)) = A(n\times n)\cdot\tilde{\chi}(|q_t|(n))$ ($A(n\times n)$ corresponds to the coefficients in Eq. (\ref{int-ql-BS-equation})). Similar methods are also adopted in solving d Lippmann-Schwinger equation for $p\bar{p}$ \cite{Kang} and $\Lambda_c\bar{\Lambda}_c$ \cite{Dai:2017fwx}.

\begin{figure}[htbp]
\centering
\subfigure[]{
\begin{minipage}[t]{0.45\linewidth}
\centering
\includegraphics[width=3.1in]{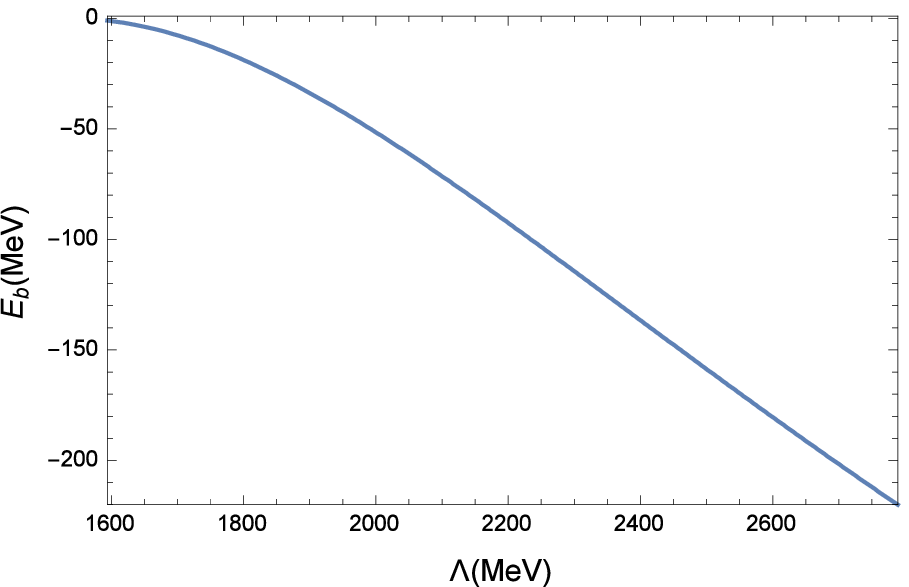}
%\caption{fig1}
\end{minipage}%
}%
\subfigure[]{
\begin{minipage}[t]{0.5\linewidth}
\centering
\includegraphics[width=3.1in]{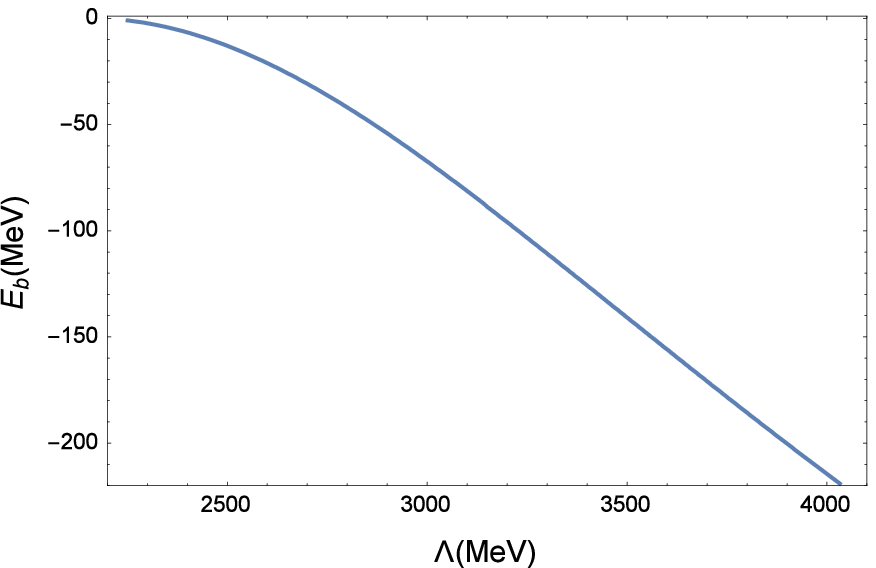}
%\caption{fig2}
\end{minipage}
}%
\centering
\caption{Relation of the cutoff $\Lambda$ and the binding energy $E_b$ with (a) the monopole form factor and (b) the dipole form factor for $I=0$.}
\label{I=0}
\end{figure}

\begin{figure}[htbp]
\centering
\subfigure[]{
\begin{minipage}[t]{0.45\linewidth}
\centering
\includegraphics[width=3.1in]{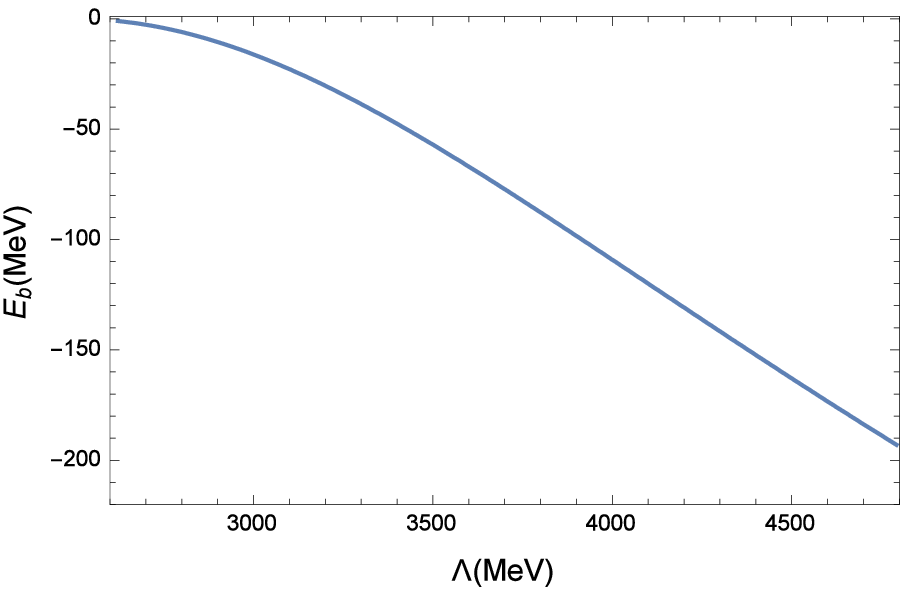}
%\caption{fig1}
\end{minipage}%
}%
\subfigure[]{
\begin{minipage}[t]{0.5\linewidth}
\centering
\includegraphics[width=3.1in]{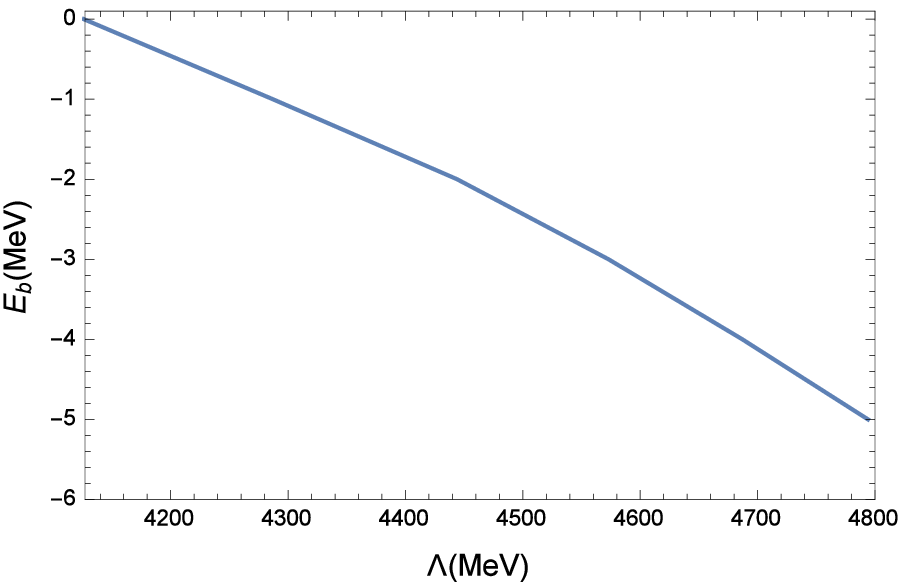}
%\caption{fig2}
\end{minipage}
}%
\centering
\caption{Relation of the cutoff $\Lambda$ and the binding energy $E_b$ with (a) the monopole form factor and (b) the dipole form factor for $I=1$.}
\label{I=1}
\end{figure}

In our calculation, we choose to work in the rest frame of the bound state in which $P = (M, 0)$. We take the averaged masses of the mesons from the PDG \cite{Tanabashi:2018oca}, $M_B=5279.41$ MeV, $M_K = 494.98$ MeV, $M_\rho=775.26$ MeV, and $M_\omega=$ 782.65 MeV. With the above preparation, we try to search for the all the possible solutions by solving the BS equation. The relation between $\Lambda$ and $E_b$ for the $B\bar{K}$ with $I=0,1$ are depicted in Fig. \ref{I=0} and Fig. \ref{I=1}, respectively.

\section{summary}
Stimulated by $X(5568)$, which is recent discover by the D0 Collaboration, we carried out a study of the interaction of $B\bar{K}$ system with isospin $I$ =0, 1 in the Bethe-Salpeter equation approach. In order to solve the BS equation, we have used the ladder approximation and the instantaneous approximation. Since the value of $\Lambda$ is near 1 GeV which is the typical scale of nonperturbative QCD interaction. Thus, if strictly considering this criterion of the value of $\Lambda$, we conclude that there do not exist isovector $B\bar{K}$ molecular state. And the $X(5568)$ cannot be the $B\bar{K}$ molecular state. The relation between $\Lambda$ and $E_b$ for the $B\bar{K}$ with $I=0,1$ are depicted in Fig. \ref{I=0} and Fig. \ref{I=1}, respectively.

%%%%%%%%%%%%%%%%%%%%%%%%%%%%%%%%%%%%%%%%%%%%%%%%%%%%%%%%%
\acknowledgments
One of the authors (Z.-Y.Zhen) thank Dr. Xian-Wei Kang for a very careful reading of this manuscript. This work was supported by National Natural Science Foundation of China (Projects No. 11275025, No. 11775024 and No.11605150) and K.C.Wong Magna Fund in Ningbo University.
%%%%%%%%%%%%%%%%%%%%%%%%%%%%%%%%%%%%%%%%%%%%%%%%%%%%%%%%

\end{document}